# A comparing study on optoelectronic properties of phototransistors based on MEH-PPV and PbS QD hybrids with bulk- and layer-heterojunction

Xiaoxian Song,[1, 2] Yating Zhang,[1, 2, a)] Ran Wang,[1, 2] Mingxuan Cao,[1, 2] Yongli Che,[1, 2] Jianlong Wang,[1, 2] Haiyan Wang,[1, 2] Lufan Jin,[1, 2] Haitao Dai,[3] Xin Ding,[1, 2] Guizhong Zhang,[1, 2] and Jianquan Yao[1, 2]

[1]*Institute of Laser & Opto-Electronics, College of Precision Instruments and Opto-electronics Engineering, Tianjin University, Tianjin 300072, China*

[2]*Key Laboratory of Opto-electronics Information Technology (Tianjin University), Ministry of Education, Tianjin 300072, China*

[3]*Tianjin Key Laboratory of Low Dimensional Materials Physics and Preparing Technology, School of Science, Tianjin University, Tianjin 300072, China*

As the responsivity (R) of a thin film photo detector is proportional to the product of the photo-induced carrier density (n) and mobility (μ) (Z. Sun, Z. Liu, J. Li, G.-a. Tai, S. Lau and F. Yan, Adv. Mater., 24, 5878, 2012), which of the types is conducive to photo detection, choosing between layer-heterojunction (LH) and bulk-heterojunction (BH) field effect phototransistors (FEpTs), is still unknown. A comparison study is performed based on an MEH-PPV and PbS QDs hybrid. They are both demonstrated as being ambipolar, with $\mu_E \approx \mu_H = 3.7$ $cm^2V^{-1}s^{-1}$ for BH-FEpTs and $\mu_H = 36$ $cm^2V^{-1}s^{-1}$, $\mu_E = 52$ $cm^2V^{-1}s^{-1}$ for LH-FEpTs. Due to the greatly improved μ and high channel order degree (α), the R of the LH-FEpTs reaches as high as $10^1$ A/W, whereas that of the BH-FEpTs is as low as $10^{-1}$ A/W. Although the large area of the BH improves the exciton separation degree (β) and n, the lack of effective transport becomes the main constraint on high responsivity. Therefore, LH-FEpTs are a better candidate, and a "three-high" principle of high α, β, and μ is required for high R values.

---

a) Author to whom correspondence should be addressed. Electronic mail: yating@tju.edu.cn.

FEpTs based on a solution-processed polymer, semiconductor quantum dots (QDs) and a hybrid of both have achieved ultrahigh performances[1-6] and are attracting more and more attention due to their great potential for applications in near infrared (NIR) detectors with low cost, flexibility, easy fabrication and easy integration[7-10]. As a center of exciton (electron-hole pairs) separation, heterojunctions play a critical role in photo detection, which converts photons into an electric current or voltage[11, 12]. Two typical heterojunctions for FEpT channel structures are BH and LH. Photo responses have been reported for both of them. Yan's group has reported BH-FEpTs based on a poly (3-hexylthiophene) (P3HT) and lead sulfide quantum dots (PbS QDs) hybrid. Under a bias of 100 V, NIR photo responsivity reaches as high a $10^4$ A/W[13]. Yan's group and Konstantato *et al.* have both reported on LH-FEpTs based on a graphene and PbS colloidal QDs hybrid with ultrahigh responsivity of $10^7$ A/W[14, 15].

Theoretically, the responsivity $R_{ph}$ of a thin film photoconductor is expressed as $R_{ph} = en\mu EW/P$, where e is the electronic charge, n is the density of photo-induced carriers per unit area, μ is the carrier mobility, E is the applied electric field, P is the incident optical power, and W is the width of the device[15]. Based on these parameters, each type has both advantages and disadvantages. BH has a larger area of exciton separation (high n) but irregular-shaped channels, where carriers transport inefficiently (low μ)[16]; LH has regular-shaped channels ( high μ), but a small area of exciton separation (low n) [14, 15].

However, which type achieves higher responsivity (~n×μ) is still undetermined. Here, both types are fabricated based on a poly [2-methoxy-5-(20-ethylhexyloxy- p-phenylenevinylene)] (MEH-PPV) and PbS QDs hybrid. Using a comparison method, the transfer characteristics and the responsivities of the two types are investigated experimentally. Because the mobility of LH-FEpTs is one order of magnitude larger than that of BH-FEpTs and the channel order degree of LH-FEpTs is larger than that of BH-FEpTs, the responsivity of LH-FEpTs is of the order of $10^1$ A/W, whereas that of BH-FepTs is ~ $10^{-1}$ A/W. Therefore, an effective carrier transport mechanism is the shortcoming of BH-FEpTs.

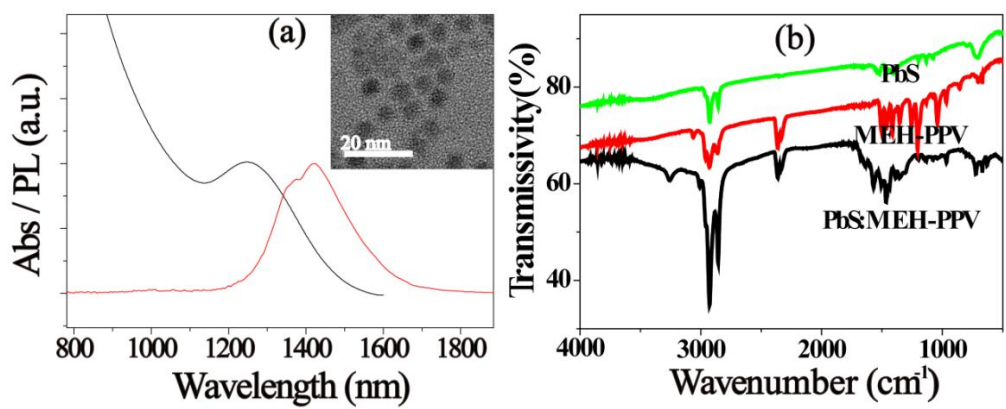

FIG. 1. (a) Absorbance and the PL spectrum of PbS QDs, the insert is TEM image of PbS QDs. (b) FTIR spectra of MEH-PPV, PbS QDs and their hybrid.

MEH-PPV was purchased from Luminescence Technology Corporation (No. LT-S931). PbS QDs was synthesized via a wet chemical method[17, 18] (for the detailed method, see supplementary material [19]). As shown in Figure 1 (a), optical characterization of the QDs, including absorption and photoluminescence (PL) spectra, was performed on a PbS QD toluene solution by using a Zolix Omni-λ300 spectrometer. The absorption peak was located at 1245 nm, while the PL peak was located at 1420 nm as excited by a 532 nm cw laser from Changchun New Industries Optoelectronics Technology Co., Ltd. According to a four-band-envelope-function formulism, the average diameter of PbS QDs was calculated as 4.8 nm[20]. The TEM image of the QDs was created using a transmission electron microscope (TEM) from FEI Co., Tecnai G2 F20, with 200 kV. Based on the TEM image, the average size of the PbS QDs was 4.8 nm, exhibiting good consistency with the diameter deduced from the absorption spectrum shown in Figure 1 (a).

The hybrid solution was prepared by combining one volume of MEH-PPV (5 mg/ml) and three volumes of PbS QDs (10 mg/ml) toluene solution. The transmission peaks of PbS QDs, MEH-PPV, and their hybrid were determined using Fourier transform infrared (FTIR) spectra (FTIR-650-spectrometer from Tianjin Gangdong sci.&tech. development Co., Ltd.), as shown in Figure 1 (b). For both PbS QDs and MEH-PPV, FTIR represents the properties of ligand (-OLA) and MEH-PPV, respectively. For the hybrid, the characteristic peaks are viewed as superpositions of the peaks from the PbS QDs and MEH-PPV individuals. Please note that there is no exchange between the ligand of PbS QDs and MEH-PPV, which ensures the consistent heterojunction of the two types.

BH-FEpTs and LH-FEpTs based on the MEH-PPV-PbS QDs hybrid were fabricated experimentally[18, 19]. The schematic diagrams of both are displayed in Figure 2 (a) and (b). The substrates used are Si n$^+$/SiO$_2$. The thickness of SiO$_2$ is 300 nm, on which an Au source and drain with a thickness of 200 nm were thermally evaporated over a shadowed mask.

The layer-heterojunction channel consists of one layer of MEH-PPV and three layers of PbS QDs, as shown in Figure 2 (a). The fabrication procedure was as follows: one drop of MEH-PPV toluene solution (5 mg/ml) was spin-casted on the substrate at a spin speed of 2000 rpm and left to dry for 15 seconds. Then, three layers of PbS QDs were deposited through a layer-by-layer film and ligand exchange process. Each layer was prepared as follows: 2% (Vol.) ethanedithiol (EDT) in acetonitrile solution was first prepared for ligand exchange. The PbS QDs toluene solution was prepared at 10 mg/mL in toluene. A drop of PbS QD layer was first deposited on the spinning substrate with a speed of 2000 rpm and left for 15 seconds to dry. Three drops of 2% EDT solution were then deposited on the rotating substrate, followed by 2 drops of acetonitrile and 2 drops of toluene.

The bulk-heterojunction channel was prepared by spin-casting four layers of the hybrid solution. Each layer was deposited at a spin speed of 2000 rpm and left for 15 seconds to dry. Then, both FEpTs were left in a vacuum overnight before electric measurement.

The SEM images of cross sections of LH-FEpTs and BH-FEpTs are given by Figure 2 (c) and (d), respectively. The structure of each device is very clearly visible. The bottom layer is Si n$^+$, on which is SiO$_2$ layer. The thickness is attributed to 300 nm. The top layer is channel layer. The thickness of such layer in LH-FEpT is 332 nm (including one layer of MEH-PPV in thickness of 93 nm and three layers of PbS QDs in thickness of 239 nm), and that of BH layer is 337 nm. Clearly, delamination presented in LH is absent in BH, due to fabrication method.

During the electric measurement, a bias voltage ($V_{SD}$) was applied over the source (ground connection) and drain electrodes by using a Keithley 2400; the channel current flowing into the drain was denoted by $I_{SD}$, which was also detected using the Keithley 2400. A gate voltage ($V_G$) was applied on the gate electrode to the ground connection by using a HP6030A, as shown in Figure 2 (b).

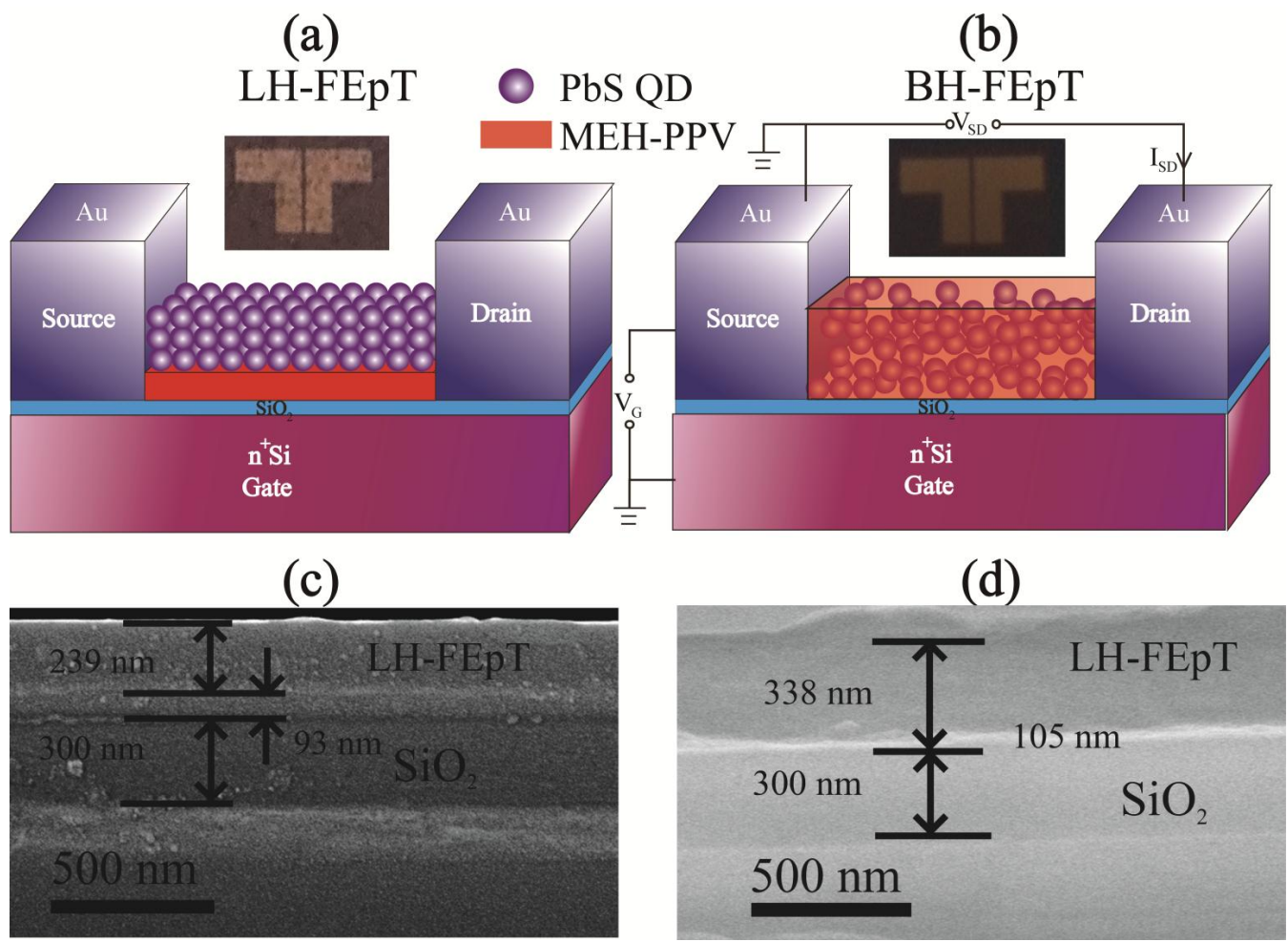

FIG. 2. Schematic diagram of the LH-FEpTs (a): one layer of MEH-PPV and three layers of PbS QDs and BH-FEpT (b): four layers of the MEH-PPV – PbS QDs hybrid solution; the inserts are top views of an LH-FEpT and a BH-FEpT, with each device being the same size (the length and width of the channel are 0.1 mm and 2.5 mm, respectively). The cross-sectional SEM image of the LH-FEpTs (c) and the BH-FEpTs (d).

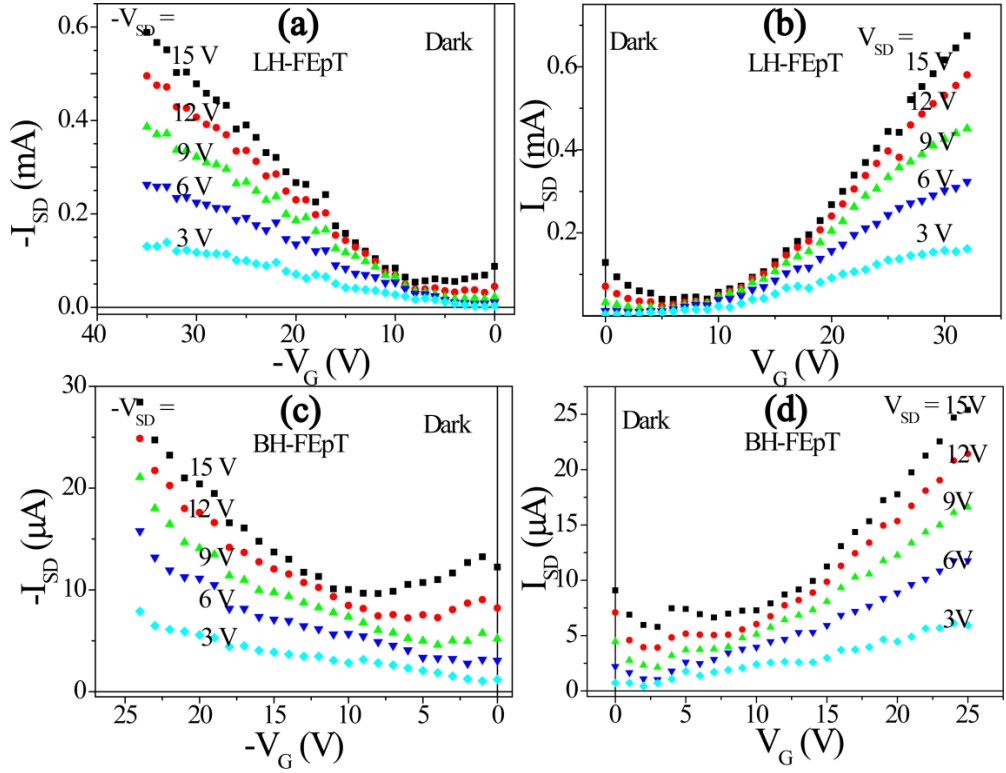

FIG. 3. Transfer characteristics ($I_{SD} \sim V_G$) of the LH-FepT are shown in the third (a) and first (b) quadrants, and those of the BH-FepT are shown in the third (c) and first (d) quadrants; both are without light irradiation and biased with ±3 V, ±6 V, ±9 V, ±12 V, and ±15 V.

Figure 3(a) and (b) present the transfer characteristics ($I_{SD} \sim V_{SD}$) of LH-FEpTs in the third and first quadrants, respectively. The transfer characteristics of BH-FEpTs are shown in the third and first quadrants of Figure 3 (c) and (d), respectively, under bias voltages of $\mp 3$ V, $\mp 6$ V, $\mp 9$ V, $\mp 12$ V and $\mp 15$ V. Both FEpTs are dipolar; although the MEH-PPV film is a p-type layer[2, 21-23], QD thin films often exhibit dipolar characteristics[7]. Similar results have been reported previously[24, 25]. The channel current of the LH-FEpTs, typically ~ 0.1 mA, is one order of magnitude larger than that of the BH-FEpTs, ~ 10μA. For a quantitative analysis, the mobility of each FEpT was calculated from the transfer curves at low-bias voltages.

Though percolation theory is always used to analyze conductivities of organic materials, the low percentages of MEH-PPV (25%) lead to the percolation mode not being applied to the hybrid [26-28]. For a QD field effect transistor (FET), the traditional theory of FET is always applied[7]. Based on that, when $V_G \gg V_{SD}$, transfer characteristics of both types of FEpTs lie in a linear region ($I_{SD} \propto V_G$) due to the constant mobility of holes (in the third quadrant) or electrons (in the first quadrant). The mobility can be expressed by[12, 20]

$$\mu = \frac{L}{WC_{ox}V_{SD}} \times \frac{I_D}{(V_G - V_T)} \tag{1}$$

where W and L are the width and length of the channel, respectively. $C_{ox}$ is the capacitance of the gate dielectric per unit area. In such a region, $I_D/(V_G - V_T) = \Delta I_D/\Delta V_G$. As a result, the slope of $\Delta I_D/\Delta V_G$ can be obtained by linear fitting on the transfer characteristic curves at a low-bias voltage, and then, the mobility can be calculated according to equation (1). With $V_{SD} = 0.3$ V, W = 2.5 mm, L = 0.1 mm, and $C_{ox} \sim 100$ pF, we determined the mobilities of holes in the third and of electrons in the first quadrants to be $\mu_H = 36$ cm$^2$V$^{-1}$s$^{-1}$ and $\mu_E = 52$ cm$^2$V$^{-1}$s$^{-1}$ for LH-FEpTs. Similarly, we also obtained the mobilities as being $\mu_H = 3.7$ cm$^2$V$^{-1}$s$^{-1}$ and $\mu_E = 3.8$ cm$^2$V$^{-1}$s$^{-1}$ for BH-FEpTs. Therefore, we conclude that the mobility of LH-FepTs is one order of magnitude higher than that of BH-FEpTs in both the third and first quadrants.

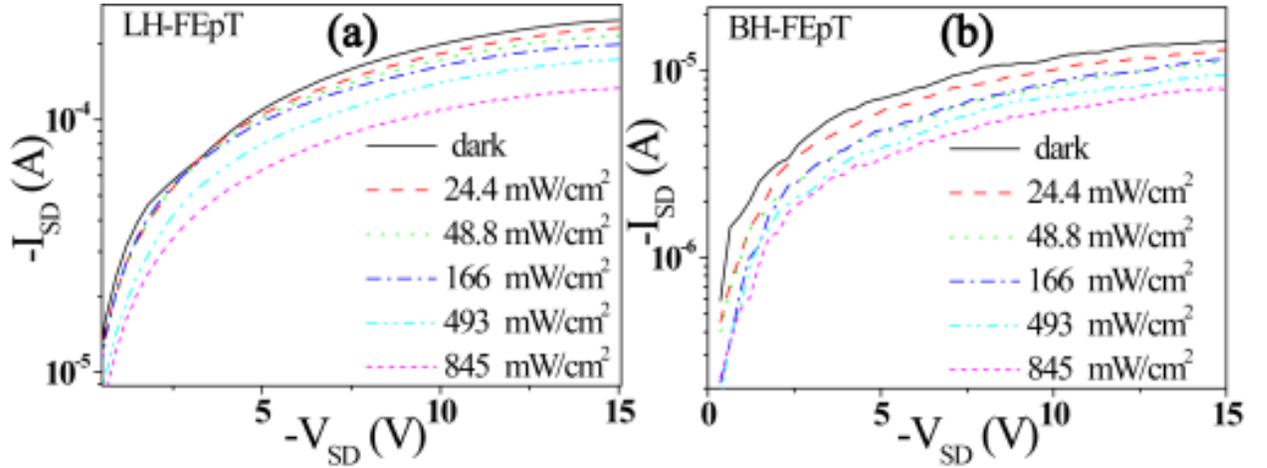

FIG. 4. Representative output characteristics ($I_{SD} \sim V_{SD}$) of LH-FEpTs (a) and BH-FEpTs in the third quadrant with a reverse gate voltage of -15 V, under light irradiations of 0 mW/cm$^2$, 24.4 mW/cm$^2$, 48.8 mW/cm$^2$, 166 mW/cm$^2$, 493 mW/cm$^2$, and 845 mW/cm$^2$.

Figure 4 (a) and (b) present the output characteristics ($I_{SD} \sim V_{SD}$) of LH-FEpTs and BH-FEpTs in the third quadrant ($V_G$ = -15 V) under light irradiations of 0 mW/cm$^2$, 24.4 mW/cm$^2$, 48.8 mW/cm$^2$, 166 mW/cm$^2$, 493 mW/cm$^2$, and 845 mW/cm$^2$.

Both types of FEpTs show the decrease of channel current with the increase in light irradiation due to the light-gate effect[13, 14, 21].

The physical mechanism of the sensing of light is described as follows. Under light illumination, photo-induced excitons are generated in the PbS QDs. Under an electric field driving force, excitons are separated into free electrons and holes in the QDs. Due to the role of gate voltage in driving carriers, holes (or electrons, depending on gate voltage) are transferred from the upper to the bottom layer where the channel was formed. Meanwhile, electrons (or holes) in equal numbers remain in the upper layers, which form the electric field in a direction opposite to that of the gate electric field. It was regarded as an additional photo-induced gate voltage, termed the "light-gate effect." Due to this effect, horizontal shifts, $\Delta V_G$, can be used with FEpTs to calibrate light irradiance ($E_e$) as [29]

$$\Delta V_G = \alpha E_e^{\beta} \tag{2}$$

where α and β are constant. According to equation (1), the increments in the channel current $\Delta I_{SD}$, resulting from the light illumination, are expressed as a function of the gate shift $\Delta V_G$:

$$\Delta I_{SD} = \frac{W}{L} C_{ox} \mu V_{SD} \Delta V_G \tag{3}$$

For one device, $WC_{ox}\mu/L$ is constant, and $\Delta I_{SD}$ is directly proportion to $V_{SD}\Delta V_G$.

However, $\Delta V_G$ is hard to measure directly. The responsivity (R) of a FEpT is always defined by $\Delta I_{SD}$ as[30]

$$R = \frac{I_{ill} - I_{Dark}}{P} = \frac{\Delta I_{SD}}{P} \tag{4}$$

where $I_{ill}$ and $I_{Dark}$ are the channel current under light illumination and in the dark, respectively, and $P = AE_e$ is the incident optical power, where A is the illumination area. Substituting equation (3) into (4), one gets

$$R = \frac{C_{ox} V_{SD} \mu}{L^2 E_e} \Delta V_G = \frac{\alpha C_{ox} V_{SD} \mu}{L^2} E_e^{\beta-1} \tag{5a}$$

$$\lg(R) = \lg\left(\frac{\alpha C_{ox} V_{SD} \mu}{L^2}\right) + (\beta - 1)\lg(E_e) \tag{5b}$$

Theoretically, lg(R) retains a linear relation with lg($E_e$), according to equation (5b). When plotting R versus $E_e$, we used a double-logarithmic system. The responsivities of LH-FEpTs in the third and first quadrants are shown in Figure 5 (a) and (b), respectively, with $V_G$ = ∓ 15 V, and $V_{SD}$ = ∓ 11 V, ∓ 18 V and ∓ 25 V. A good linear dependence of lg($E_e$) on lg(R) is observed. By fitting the data using equation (5b), β was obtained to be -0.18 and 0.026 and α was found to be $5.5 \times 10^4$ and $5.4 \times 10^4$ Vcm$^2$/W in the third and first quadrants, respectively. For BH-FEpTs, a good linear dependence of lg($E_e$) on lg(R) is also observed, as seen in Figure 5 (c) and (d). By fitting the data using equation (5b), β was obtained as 0.16 and 0.086 and α was observed to be $8.1 \times 10^3$ and $6.23 \times 10^4$ Vcm$^2$/W in the third and first quadrants, respectively.

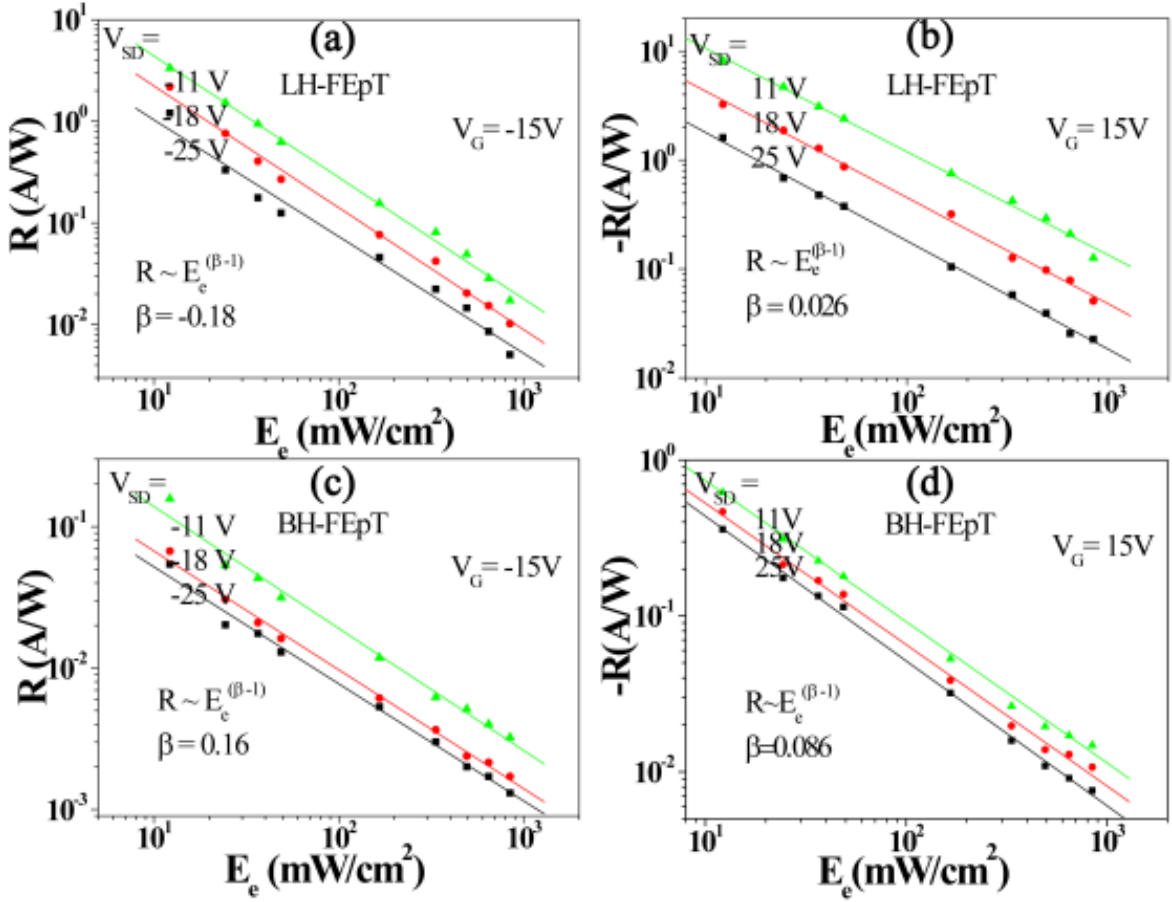

FIG. 5. Photo responsivity (R) versus light irradiation ($E_e$) of LH-FEpTs in the third (a), first (b) quadrants and of BH-FEpTs in the third (c) and first (d) quadrants, respectively.

Next, we will discuss the contribution to and influence of α, β, μ and $\alpha\mu C_{ox}/L^2$ on responsivity based on the sensing mechanism of FEpTs. Equation (5b) shows that the light response of a FEpT is related to two factors: β and (α×μ). (β-1) represents the slope of lg(R) on lg($E_e$), and (α×μ)$C_{ox}/L^2$ corresponds to the lg(R) intercept when lg($E_e$) tends to zero.

According to equation (2), β represents the photo-induced charge separation degree, while α corresponds to the order degree of the channel formed by these separated charges.

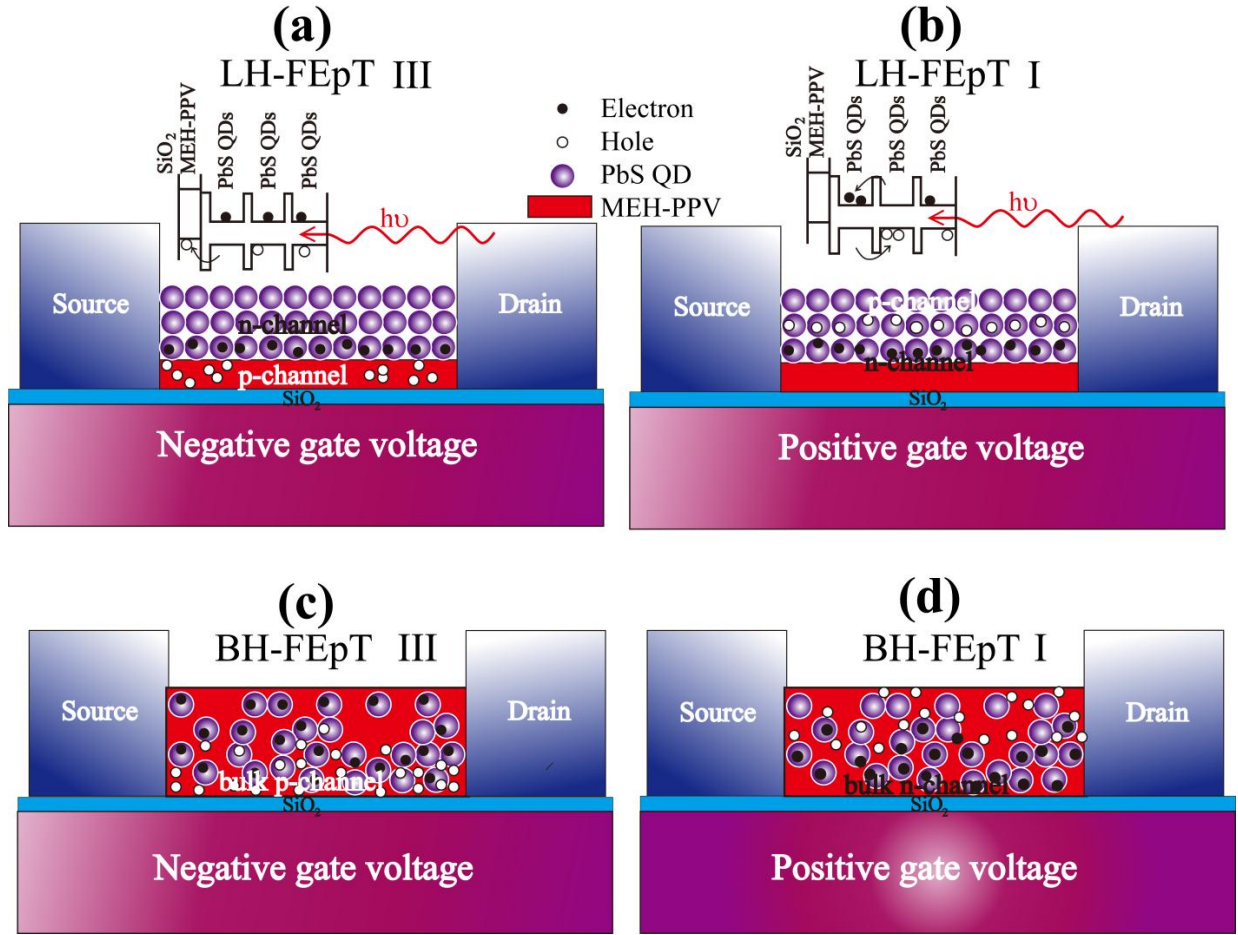

FIG. 6. Schematic diagram of the carrier transfer and formed channel for LH-FEpTs in the third (a) and first (b) quadrants and for BH-FEpTs in the third (c) and first (d) quadrants, respectively. Insert: a schematic energy transfer diagram of LH-FEpTs on an $n^+$ Si / $SiO_2$ substrate with negative (a) and positive (b) gate voltage.

In LH-FEpTs, when a negative gate voltage is applied, as in Figure 6 (a), photo-induced holes are transferred from the QDs to MEH-PPV. As a consequence, holes collected in the MEH-PPV and the bottom layer of PbS QDs where a p-channel formed. Meanwhile, electrons that remained in PbS QDs formed an additional gate voltage, representing the light-gate effect. Due to the small interfacial area of LH, the separation degree was rather low, with a value of -0.18. Then, the holes were transported either in the MEH-PPV or in the bottom layer of QDs. Due to this disorder, the mobility of MEH-PPV was of the order of $10^{-2} \sim 10^{-6}$ $cm^2V^{-1}s^{-1}$ [27, 31], while that of QDs FET was ~ 10 $cm^2V^{-1}s^{-1}$ [7], indicating that holes in the QDs layer are transported more effectively than in MEH-PPV. Thus, the measured mobility (36 cm$^2$V$^{-1}$s$^{-1}$) is mainly contributed by holes in the QDs, which is also a reasonable value for mobility in QD FETs according to a prior report [32].

When the positive gate voltage was applied, the carrier transferred to the homojunction interface instead of the heterojunction interface due to the level mismatch between QDs and MEH-PPV. As the alignment of the energy level of the QD layers occurred, as in Figure 6 (b), the separation degree improved to β = 0.026. After that, an n--channel formed in the bottom layers of the PbS QDs. It has been reported that the pristine QD FEpTs also exhibited a photo response[31]. For a layered structure, the n-channel order degree is equivalent to the p-channel; then, a very close value of α is obtained. As a consequence, a larger αμC$_{ox}$/L$^2$ and a higher responsivity (10$^1$ A/W) are achieved by LH-FEpTs in the positive gate operating region due to the high μ, high β and equivalent α.

|  | Transport μ (cm$^2$V$^{-1}$s$^{-1}$) | Channel order α (Vcm$^2$/W) | Transfer β | Transport+order αμC$_{ox}$/L$^2$ (A/WV) |
|---|---|---|---|---|
| LH-FEpT in III | 36 | 5.5×10$^4$ | -0.18 | 2.0 |
| LH-FEpT in I | 52 | 5.4×10$^4$ | 0.026 | 2.8 |
| BH-FEpT in III | 3.7 | 8.1×10$^3$ | 0.16 | 0.03 |
| BH-FEpT in I | 3.8 | 6.2×10$^4$ | 0.086 | 0.0623 |

Table 1. Value table of four parameters (μ, α, β and αμC$_{ox}$/L$^2$) for LH-FEpT and BH-FEpT in third (III) and first (I) quadrants.

In BH-FEpTs, due to the large interfacial area of the heterojunction, the value of β is higher than that of LH-FEpTs. The difference in values of β in the positive and negative gate operating regions is attributed to up-down asymmetry. The mobility is reduced greatly by the scattering process of space charges. Due to the up-down asymmetry of BH, when the opposite gate voltage was applied, the channel order degrees (α) exhibited different values. The larger the interfacial area was, the larger the separation degree (β) of photon excitons was and the lower the channel order (α) was. Obviously, the α and β values of BH-FEpTs show opposite trends. As a result, in BH-FepTs, a higher αμC$_{ox}$/L$^2$ is obtained in a positive gate operating region.

Comparing the two types of FEpTs, LH-FEpTs shows higher μ and higher (α×μ) than BH-FEpTs. Although the larger area of heterojunction results in a high β and a large concentration of carriers in BH-FEpTs, the low μ and low α of the bulk channel constrain high responsivity. Finally, an ideal FEpT with high responsivity can be obtained via a "three-high" principle, which involves high α, high β, and high μ.

In summary, we investigated two types of FEpT NIR detectors based on a MEH-PPV and PbS QDs hybrid, LH-FEpTs and BH-FEpTs. Bipolar characteristics were shown by both. The $\mu_H \approx \mu_E$ in BH-FEpTs is 3.7 cm$^2$V$^{-1}$s$^{-1}$, and $\mu_H$ = 36 cm$^2$V$^{-1}$s$^{-1}$, $\mu_E$ = 52 cm$^2$V$^{-1}$s$^{-1}$ in LH-FEpTs. Due to the greatly promoted mobility and highly ordered channels, the responsivity of LH-FEpTs can reach as high as $10^1$ A/W, whereas that of BH-FEpTs can be as low as $10^{-1}$ A/W. Although the large area of BHs shows a high separation degree of photo-excitons, the lack of an effective transport mechanism becomes the main restraining factor. Therefore, LH-FEpTs are a better candidate for a NIR photo detector. Finally, an ideal FEpT with high responsivity is obtained via a "three-high" principle with high α, high β, and high μ.

This work is supported by the National Natural Science Foundation of China (Grant No. 61271066) and the Foundation of Independent Innovation of Tianjin University (Grant No. 60302070).